\documentstyle[prl,aps,multicol,epsf]{revtex}
\renewcommand{\narrowtext}{\begin{multicols}{2} \global\columnwidth20.5pc}
\renewcommand{\widetext}{\end{multicols} \global\columnwidth42.5pc} 

\begin{document}

\newcommand{\be}{\begin{equation}}
\newcommand{\ee}{\end{equation}}
\newcommand{\bea}{\begin{eqnarray}}
\newcommand{\eea}{\end{eqnarray}}
\newcommand{\nt}{\narrowtext}
\newcommand{\wt}{\widetext}

\title{Elusive gauge-invariant fermion propagator in QED-like effective theories: round II}

\author{D. V. Khveshchenko}

\address{Department of Physics and Astronomy, University of North
Carolina, Chapel Hill, NC 27599}
\maketitle

\begin{abstract}
We comment on the recent attempt by M. Franz et al \cite{Vafek2}
to further justify their earlier calculation of the gauge-invariant electron propagator
in the context of the $QED_3$ theory of the pseudogap phase in cuprates \cite{Vafek1}.
First, we use the method of "$reductio$ $ad$ $absurdum$" to demonstrate
the inconsistency of the argument offered in \cite{Vafek2}
and then present a direct calculation of the disputed fermion amplitudes,
thus unequivocally proving that the previously proposed form of the electron propagator:
1) does exhibit a negative anomalous dimension, as pointed out in \cite{DVK1}; 
2) is different from the so-called Brown's function whose anomalous 
dimension turns out to be positive when computed in a covariant gauge.
Lastly, we conjecture that the true physical electron propagator 
(whose exact form still remains to be found)
may feature a "super-Luttinger" behavior characterized by a faster than a power-law
decay $G_{phys}(x)\propto\exp(-const\ln^2(\Lambda |x|))$.
\end{abstract}

\vspace{1.0cm}
In their recent note \cite{Vafek2}, 
M. Franz et al continued to advocate the naive "stringy ansatz" 
with the inserted Wilson line-like factor taken along the straight path
$\Gamma$ between the end points $x$ and $y$ 
\be
G_0(x-y)=<0|\psi(x)\exp(-i\int_\Gamma A_\mu(z)dz^\mu){\overline \psi}(y)|0> 
\ee
as a viable candidate for the gauge invariant propagator of physical electrons 
in the $QED_3$ theory of the pseudogap phase in cuprates \cite{Vafek1}.
In this theory, the electron operators are related by virtue of a singular gauge 
transformation $\Psi(x)=e^{i\theta(x,\infty |A)}\psi(x)$
to the $N$-flavored Lagrangian fermion variables (in the theory of the pseudogap phase of  
Ref.\cite{Vafek2} $N=2$) which are governed by the $QED_3$ action
\be
S[\psi,{\overline \psi},{A}]=
\int {\bf dz}[\sum_{f=1}^N{\overline \psi}_f
(i{\hat \gamma}_\mu \partial_\mu+{\hat \gamma}_\mu A_\mu)
\psi_f+{1\over 4g^2}(\partial_\mu A_\nu - \partial_\nu A_\mu)^2]
\ee 
In the physically interesting strong coupling regime of momenta $q\lesssim \Lambda=Ng^2$
the effect of fermion polarization is described by the renormalized gauge field propagator
\be
{D}_{\mu\nu}(q)=<0|A_\mu(q)A_\nu(-q)|0>={N\over 8}{\sqrt {-q^2}}[\delta^{\mu\nu}+(\lambda-1)
{q^\mu q^\nu\over q^2}]
\ee
which is parametrized by $\lambda$ in the class of covariant (generalized Lorentz) gauges. 
 
In particular, the authors of \cite{Vafek1} claimed that the amplitude (1)
exhibits a physically sensible behavior characterized by a positive anomalous dimension.
This conclusion was drawn solely on the basis of a seemingly
convincing argument that the amplitude (1) can be identified with another 
function known as the Brown's construct \cite{Brown}
\be
G_1(x-y)={<0|\psi(x){\overline \psi}(y)|0>\over <0|\exp(i\int_\Gamma A_\mu(z)dz^\mu)|0>}
\ee
whose anomalous dimension can be deduced rather straightforwardly 
from the ratio between the wave function renormalization factor 
determining the anomalous dimension of the ordinary
(gauge variant) fermion propagator \cite{Nash}
 
\be
Z_\psi(x-y)={<0|\psi(x){\overline \psi}(y)|0>\over G(x-y)} 
\propto (\Lambda |x-y|)^{(4/3\pi^2N)(2-3\lambda)}
\ee
(hereafter $G(x-y)$ stands for the bare propagator) and the Gaussian average of the Wilson line
\be
<\exp(i\int_\Gamma A_\mu dz_\mu)>=
\exp[-{1\over 2}\int_\Gamma dz_1^\mu \int_\Gamma dz_2^\nu D_{\mu\nu}(z_1-z_2)]\propto
(\Lambda |x-y|)^{(4/\pi^2N)(2-\lambda)},
\ee
thus resulting in the overall $positive$ anomalous dimension 
\cite{Vafek1}
\be
\eta^{3D}_1={16\over 3\pi^2N}
\ee
which is free of the gauge parameter $\lambda$, thus creating the impression
that Eq.(4) represents a truly gauge invariant function. If this were the case,
then one would indeed be able to identify Eqs.(1) and (4) by simply noticing  
that the two functions appear to coincide in the axial (Fock-Shwinger) 
gauge $(x-y)_\mu A_\mu(z)=0$ where the Wilson line factor equals unity.
 
In their recent follow-up note \cite{Vafek2}, M. Franz et al
attempted to further support these conclusions by invoking a formal 
textbook argument \cite{Zinn-Justin} which allows one to switch 
between different gauges when computing a gauge invariant 
quantity such as, e.g., partition sum. In fact, it is a confidence in the (already
established) gauge invariance of an amplitude 
in question that makes this argument a meaningful
statement, while, in the absence of such, this whole argument becomes largely irrelevant.

In our earlier note \cite{DVK2}, we argued that
the argument of Ref.\cite{Vafek1} does have a caveat, for the Brown's function
defined as a $ratio$ of the two different amplitudes can not be truly
gauge invariant, for the gauge fields in the numerator and 
denominator of Eq.(4) may transform totally independently of one another 
($A^{1,2}_\mu\to A^{1,2}_\mu+\partial_\mu f^{1,2}$),
thus resulting in the overall phase factor $e^{if^1(x)-if^2(x)-if^1(y)+if^2(y)}$ 
which only vanishes for $f^1(x)=f^2(x)$.

Albeit not exactly gauge invariant, the Brown's function is, nevertheless, independent of 
the $\it gauge$ $\it parameter$ $\lambda$ in the particular  
class of the covariant gauges, as manifested by Eq.(7). What invalidates the 
attempt to identify Eqs.(1) and (4) despite their coinciding with each other
in the axial gauge is the fact that the latter gauge is not a covariant one.
Therefore, the surrogate Eq.(4) can not substitute for the amplitude (1) whose dimension 
was previously found to be $negative$ \cite{DVK1}
\be
\eta^{3D}_0=-{32\over 3\pi^2N},
\ee
thus making Eq.(1) a rather poor candidate for the physical electron propagator. 

In fact, if the conjectured identity between Eqs.(1) and (4) 
proved to be true, it would also hold for any function defined as 
\be
G_\xi(x-y)={<0|\psi(x)\exp(i(\xi-1)\int_\Gamma A_\mu(z)dz_\mu)|0>
{\overline \psi}(y)|0>\over <0|\exp(i\xi\int_\Gamma A_\mu(z)dz_\mu)|0>},
\ee
because: 1) Eq.(9) appears to be seemingly gauge invariant to the same extent as 
Eq.(4) (that is, provided that one uses the $same$ gauge transformation in both numerator and denominator); 
2) it also coincides with both Eqs.(1) and (4) when computed in
the axial gauge (observe that Eqs.(1) and (4)
correspond to $\xi=0$ and $\xi=1$, respectively). 
The parameter $\xi$ should not be confused with the gauge parameter $\lambda$ 
(for a reader's convenience, we point out 
that in Refs.\cite{Vafek2,Vafek1} the gauge parameter is denoted
as $\xi$, while our parameter $\xi$ has no counterpart).

It can be readily seen, however, that the alleged $\xi$-independence of Eq.(9) would, in fact, 
be too much to wish for. To this end, we represent Eq.(9) as a functional average
(hereafter denoted by brackets $<\dots>$) 
\be
G_\xi(x-y)={<{\cal G}(x,y|A)e^{i(\xi-1)\theta(x,y|A)}>\over
<e^{i\xi\theta(x,y|A)}>},
\ee
where $\theta(x,y|A)=\int_\Gamma A_\mu(z)dz^\mu$ and 
${\cal G}(x,y|A)=1/(i{\hat \partial}+{\hat A})$
stands for the inverse Dirac operator, 
over different gauge field configurations with the weight determined by the effective action 
\be
S_{eff}[{A}]={1\over 2}
\int {\bf dx}\int {\bf dy} A_\mu(x)D^{-1}_{\mu\nu}(x-y)A_\nu(y)
\ee
where, in accord with all the previous work on the subject, 
we neglect higher order (non-Gaussian) corrections produced by fermion polarization,
thereby focusing on the leading terms in the $1/N$ expansion.

It can be readily seen that, if Eq.(9) were indeed independent of $\xi$, one would be able
to establish a number of clearly improbable identities, including
\be
<{\cal G}(x,y|A)e^{i(\xi-1)\theta(x,y|A)}>=
<{\cal G}(x,y|A)e^{-i(\xi+1)\theta(x,y|A)}>~~~~~~(WRONG)
\ee
which stems from equating (9) to its value obtained for $-\xi$
and noticing that in a covariant gauge 
(and in the leading $1/N$ approximation) one has $<e^{i\xi\theta(x,y|A)}>=<e^{-i\xi\theta(x,y|A)}>$ (see (6)).

In particular, if valid, Eq.(12) would have implied that
the ordinary (gauge-variant) fermion propagator 
$<0|\psi(x){\overline \psi}(y)|0>=<{\cal G}(x,y|A)>$ coincides with the amplitude
$<{\cal G}(x,y|A)e^{-2i\theta(x,y|A)}>$ not only in the axial (where
it does hold) but also in an arbitrary, including any covariant, gauge.

And it gets even better: by differentiating Eq.(9) with 
respect to $\xi$ and putting $\xi=0$ one finds that the alleged
$\xi$-independence (hence, the requirement $dG_\xi(x-y)/d\xi|_{\xi=0}=0$) 
imposes the condition   
\be
i<{\cal G}(x,y|A)\theta(x,y|A)e^{-i\theta(x,y|A)}>=0~~~~~(WRONG)
\ee
The validity of (13) can be ascertained by simply computing this amplitude,
for which purpose it would be most convenient to use the exact quantum mechanical
(i.e., single-particle) path integral representation 
of the inverse Dirac operator ${\cal G}(x,y|A)$ \cite{Skachkov,Stefanis,DVK1} (also, see \cite{us}
for an asymptotically exact non-perturbative calculation of the gauge-invariant 
fermion propagator in the case of a static gauge field which one encounters in such problems
as the effect of vortex disorder on quasiparticle properties of $d$-wave superconductors 
or that of dislocations in layered graphite).
However, here we choose a more traditional approach, for we recognize that 
the more powerful method devised in \cite{Skachkov,Stefanis}
and further advanced in \cite{DVK1,us} may still be somewhat less familiar.

In fact, it suffices to compute (13) to first order in $1/N$ by expanding both 
the exponential factor $e^{-i\theta(x,y|A)}=
1-i\theta(x,y|A)+\dots$ and the inverse Dirac operator $1/(i{\hat \partial}+{\hat A})=
1/i{\hat \partial}-(1/i{\hat \partial}){\hat A}(1/i{\hat \partial})+\dots$
in powers of $A_\mu(z)$ (apparently, no cancellation can  
possibly occur between different $1/N$ order terms)
$$
i<{\cal G}(x,y|A)\theta(x,y|A)e^{-i\theta(x,y|A)}>=
G(x-y)\int_\Gamma d{z}_1^\mu \int_\Gamma dz_2^\nu D_{\mu\nu}(z_1-z_2)-
$$
\be
-\int d{\bf z}_1G(x-z_1)\gamma_\mu G(z_1-y)\int_\Gamma dz_2^\nu
D_{\mu\nu}(z_1-z_2)
=-{16\over \pi^2N}G(x-y)\ln(\Lambda|x-y|)\neq 0~~~~~~(Q.E.D.)
\ee
where we used the following Eqs.(17) and (18) (see below)
after scaling out the factors $(1-2\xi)$ and $(1-\xi)$,
respectively. Taken at its face value, the result (14) implies that $G_0(x-y)\neq G_1(x-y)$,
in complete accord with the proverbial moral: 
"the average of a product is not necessarily equal to a product of the averages".  

In fact, the sought-after identity between the different functions $G_\xi(x)$
can approximately hold only in the case of massive fermions and only in the vicinity
of the mass shell ($|p^2-m^2|\ll m^2$), 
resulting in their common (rather trivial) long-distance behavior
$G_\xi(x)\propto e^{-m|x|}$. In this regime, the leading functional dependence on the gauge  
field reduces to the eikonal phase factor ${\cal G}(x,y|A)\sim e^{i\theta(x,y|A)}$ \cite{DVK1,Skachkov,Stefanis},
and it is precisely this property that the Brown's function was designed to make use of 
in the first place \cite{Brown}.

However, in the massless case such a regime 
is absent altogether, and the long-distance behavior
is solely controlled by the ultra-violet (UV) anomalous dimensions  
(see \cite{DVK1,DVK2} for a more extensive discussion of this potentially confusing issue).

Having completed our proof, we now present a direct calculation 
of $G_\xi(x)$ for an arbitrary $\xi$ in a generic covariant $\lambda$-gauge.
After expanding (9) to the second order in $A_\mu(z)$, 
we find three different kinds of correction terms
$$
G_\xi(x-y)-G(x-y)=-\int{\bf dz}_1{\bf dz}_2
<G(x-z_1){\hat A}(z_1)G(z_1-z_2){\hat A}(z_2)G(z_2-y)>+
$$
\be
+{\xi^2-(1-\xi)^2\over 2}G(x-y)<(\theta(x,y|A))^2>+
(1-\xi)\int{\bf dz}<G(x-z){\hat A}(z)G(z-y)\theta(x,y|A)>
\ee
First of these terms corresponds to
the lowest order self-energy correction to the ordinary Green function 
\be
\delta_1G_\xi(x-y)=-\int d{\bf z}_1d{\bf z}_2 G(x-z_1)\gamma_\mu 
D_{\mu\nu}(z_1-z_2)\gamma_\nu G(z_2-y)=
G(x-y){4\over 3\pi^2N}(2-3\lambda)\ln(\Lambda |x-y|)
\ee
thus reproducing the anomalous dimension associated with the wave function
renormalization factor (5).

The second type of corrections originates from the expansion of the Wilson lines
inserted to the numerator and denominator of Eq.(9)
\be
\delta_2G_\xi(x-y)={\xi^2-(1-\xi)^2\over 2}G(x-y)\int_\Gamma dz_1^\mu \int_\Gamma dz_2^\nu 
D_{\mu\nu}(z_1-z_2)=G(x-y){4\over \pi^2N}(1-2\xi)(2-\lambda)\ln(\Lambda |x-y|)
\ee
which, too, can be readily exponentiated and yields the anomalous dimension
equal to that given by Eq.(6) with the extra factor $(1-2\xi)$. 

Lastly, there is a mixed term which stems from the first 
order expansion of both $G(x,y|A)$ and the Wilson line. Unlike Eqs.(16) and (17), 
this contribution has not been discussed in the previous studies, 
and, therefore, we spell its evaluation out in all the details
$$
\delta_3G_\xi(x-y)=(1-\xi)\int d{\bf z}_1G(x-z_1)\gamma_\mu G(z_1-y)\int_\Gamma dz_2^\nu
D_{\mu\nu}(z_1-z_2)\approx
$$ 
$$
\approx(1-\xi)\int d{\bf z}_1[G(x-y)\gamma_\mu G(z_1-y)+G(x-z_1)\gamma_\mu G(x-y)]\int_\Gamma dz_2^\nu
D_{\mu\nu}(z_1-z_2)=
$$
$$
={2\over \pi^3N}(1-\xi)G(x-y)\gamma_\mu\int_\Gamma dz_2^\nu\int{\bf dz}_1
{(z_1^\nu-y^\nu)\over |z_1-y|^3}{1\over |z_1-z_2|^2}
[\lambda\delta_{\mu\nu}+2(1-\lambda){(z_1^\mu-z_2^\mu)(z_1^\nu-z_2^\nu)\over
|z_1-z_2|^2}]
=
$$
\be
=G(x-y){8\over \pi^2N}(1-\xi)\lambda \ln(\Lambda |x-y|)
\ee
Eq.(18) was obtained with logarithmic accuracy, and 
in the course of this calculation we used the 3D real space propagators
\be
G(x)={1\over 4\pi}{{\hat x}\over x^3}, ~~~~~D_{\mu\nu}(x)={4\over \pi^2N|x|^2}
[\lambda\delta_{\mu\nu}+2(1-\lambda){x^\mu x^\nu\over |x|^2}]
\ee
and the following D-dimensional integrals 
\be
\int {\bf dx}{x^\alpha\over |x|^D|x-y|^{D-1}}={2\pi^{D/2}\over \Gamma(D/2)}
{y^\alpha\over |y|^{D-1}} 
\ee
and
\be
\int {\bf dx}{x^\alpha(x^\beta-y^\beta)(x^\gamma-y^\gamma)
\over |x|^D|x-y|^{D+1}}={2\pi^{D/2}\over 3(D-1)\Gamma(D/2)}
[{2y^\alpha\delta^{\beta\gamma}-y^\beta\delta^{\alpha\gamma}-y^\gamma\delta^{\alpha\beta}
\over |y|^{D-1}} 
+
(D-1){y^\alpha y^\beta y^\gamma\over |y|^{D+1}}]
\ee
Combining Eqs.(16), (17), and (18) together we finally obtain the anomalous dimension
of $G_\xi(x)$ in the form 
\be
\eta^{3D}_\xi={16\over 3\pi^2N}(3\xi-2)
\ee
Thus, despite the fact that the gauge parameter $\lambda$ cancels out, as expected,
the functions $G_\xi(x)$ are starkly different for different $\xi$.
In particular, Eqs.(7) and (8) follow from (22) for $\xi=1$ and $\xi=0$, respectively.

Anticipating possible objections that the above results might have been different,
should we have resorted to the dimensional regularization of our divergent integrals,
we have also confirmed that in $D\to 3$ dimensions Eqs.(16), (17), and (18) are all proportional 
to $\int_\Gamma dz^\mu(z-y)^\mu/|z-y|^{D-1}=(\Lambda |x-y|)^{3-D}/(3-D)\to\ln(\Lambda|x-y|)$.

For the sake of completeness, here we also present the results for the weak coupling $QED_4$
which demonstrate that the situation in 3D is not at all exceptional.
Instead of Eqs.(16),(17), and (18) we now get 
\be
\delta_1G_\xi(x-y)=-G(x-y){g^2\over 8\pi^2}\lambda\ln(\Lambda |x-y|)
\ee
\be
\delta_2G_\xi(x-y)=G(x-y){g^2\over 8\pi^2}(1-2\xi)(3-\lambda)\ln(\Lambda |x-y|)
\ee
\be
\delta_3G_\xi(x-y)=G(x-y){g^2\over 4\pi^2}(1-\xi)\lambda \ln(\Lambda |x-y|)
\ee
Instead of Eq.(19) we use
\be
G(x)={1\over 2\pi^2}{{\hat x}\over x^4}, ~~~~~D_{\mu\nu}(x)={g^2\over 4\pi^2|x|^2}
[{1+\lambda\over 2}\delta_{\mu\nu}+(1-\lambda){x^\mu x^\nu\over |x|^2}]
\ee
and because of a different power-counting the 4D calculation involves the integrals 
\be
\int {\bf dx}{x^\alpha\over |x|^D|x-y|^{D-2}}={\pi^{D/2}\over \Gamma(D/2)}
{y^\alpha\over |y|^{D-2}} 
\ee
and
\be
\int {\bf dx}{x^\alpha(x^\beta-y^\beta)(x^\gamma-y^\gamma)
\over |x|^D|x-y|^{D}}={\pi^{D/2}\over 2(D-2)\Gamma(D/2)}
[{y^\alpha\delta^{\beta\gamma}-y^\beta\delta^{\alpha\gamma}-y^\gamma\delta^{\alpha\beta}
\over |y|^{D-2}} 
+
(D-2){y^\alpha y^\beta y^\gamma\over |y|^{D}}]
\ee
Combining Eqs. (23), (24), and (25) together we finally obtain the total anomalous dimension
of $G_\xi(x)$ in the 3D case 
\be
\eta^{4D}_\xi={3g^2\over 8\pi^2}(2\xi-1)
\ee 
Interestingly enough, the values of Eq.(29) obtained for $\xi=0$ and $\xi=1$ differ only in their sign.

We emphasize that the $\it positive$ values of $\eta^{3D,4D}_1$ pertaining to 
the (not exactly gauge-invariant) function $G_1(x)$ computed in the covariant gauge does not
fix the problem with the $\it negative$ anomalous dimension $\eta^{3D,4D}_0$ of 
the conjectured form of the physical electron propagator which
is given by the (exactly gauge-invariant) function $G_0(x)$.
In light of these findings, in Refs.\cite{DVK1,DVK2} we suggested to actively explore 
alternate proposals for the physical electron propagator
which, due to the nature of the relationship between the electrons 
and the Lagrangian fermions implemented through a singular
gauge transformation, can only be given by a 
$single$ gauge field average, not a $ratio$ of such.

To this end, in \cite{DVK1,DVK2} we discussed the "dressed charge" propagator constructed 
in the context of the conventional (i.e., massive) $QED_4$ \cite{Lavelle}. 
The authors of Ref.\cite{Lavelle} showed that the Fourier
transform of this gauge invariant amplitude 
\be
G_v(x-y)=<0|\psi(x)\exp[i{\delta^{\mu\nu}-({u}^\mu+v^\mu)({u}^\nu-v^\nu)\over 
\partial_\alpha^2-({u}_\alpha\partial_\alpha)^2+(v_\alpha\partial_\alpha)^2}\partial_\mu A_\nu]
{\overline \psi}(y)|0>
\ee
where $u^\mu=(1,{\vec 0})$ and $v^\mu=(0,{\vec v})$
features a simple pole-like behavior 
at the single point $p^\mu=m(1,{\vec v})/{\sqrt {1-v^2}}$ on the mass shell  
which corresponds to a charge moving with a velocity $\vec v$.
The UV anomalous dimension of $G_v(x)$ computed in \cite{Lavelle} 
\be
\eta^{4D}_v=-{g^2\over 8\pi^2}[3+2{1\over v}\ln{1-v\over 1+v}]
\ee 
remains $positive$ for any $v$ and increases logarithmically towards infinitely high values as $v\to 1$.

Albeit not being immediately applicable to the case $m=0$, the calculation carried out
in Ref.\cite{Lavelle} suggests that, as one proceeds beyond the first order, 
the logarithmic growth of (31) gets cut off at $max(|v-1|, 1/\Lambda|x|)$.
Therefore, it is not totally inconceivable
that the massless counterpart of Eq.(30) may exhibit a faster than a power-law decay
\be
G_{phys}(x)\propto\exp(-const\ln^2(\Lambda|x|)),
\ee
where, depending on the dimension, the constant is proportional to either $1/N$ or $g^2$,
thus placing the effective $QED$-like theories of condensed matter systems
into the class of "super-Luttinger" models,
alongside the 1D metals with unscreened Coulombic interactions where $G_{phys}(x)\propto\exp(-const\ln^{3/2}(\Lambda|x|))$.

In light of such a possibility, the previous attempts to discover 
the Luttinger-like behavior in the pseudogap phase of the cuprates \cite{ARPES} 
may need to be prepared to handle a potentially
much stronger suppression of the physical amplitudes in order to reconcile
the predictions of the $QED_3$ theory with the available photoemission, tunneling, and other data.

We conclude by stressing that, arguably, the problem of constructing the gauge 
invariant fermion propagator in massless $QED$, thus far, 
has received a lesser attention than it deserves.
The intricacy of the related calculations indicates that   
this problem really needs to be settled before one can start drawing solid
(instead of wishful) conclusions about the true behavior in the $QED$-like as
well as other gauge field models, including non-abelian and discrete symmetry (say, $Z_n$) ones. 
A successful completion of this task is likely to require some new ideas 
and/or potentially cumbersome calculations. 
Quoting from yet another textbook, in addition to the already 
mentioned ones \cite{Brown,Zinn-Justin}, simply wouldn't do.

The author acknowledges valuable email communications with V. Gusynin and I. Herbut. 
This research was supported by the NSF under Grant No. DMR-0071362.

\end{document}